\documentstyle[aps,epsf]{revtex} 
\begin{document}
\title{{\hfill} {\sl Submitted}\\ 
Envelope solitons induced by high-order effects \\
of light-plasma interaction}
\author{P. K. Shukla, R. Fedele\cite{a)}, M. Onorato\cite{b)},
N. L. Tsintsadze\cite{c)} } \address{Institut f\"{u}r
Theoretische Physik IV, Fakult\"{a}t f\"{u}r Physik und
Astronomie,\\ Ruhr-Universit\"{a}t Bochum, D-44780 Bochum,
Germany}
\maketitle
\begin{abstract}
The nonlinear coupling between the light beams and
non-resonant ion density perturbations in a plasma is
considered, taking into account the relativistic particle
mass increase and the light beam ponderomotive force. A
pair of equations comprising a nonlinear Schr\"{o}dinger
equation for the light beams and a driven (by the light
beam pressure) ion-acoustic wave response is derived. It
is shown that the stationary solutions of the nonlinear
equations can be  represented in the form of a bright and
dark/gray soliton for one-dimensional problem. We have
also present a numerical analysis which shows that our
bright soliton solutions are stable exclusively for the
values of the parameters compatible with of our theory.
\end{abstract}

\noindent
\pacs{PACS Number (s): 52.35.Mw, 52.40.Db, 52.60.+h, 52.35.Sb}
%

\section{Introduction}
Recently, the investigations concerning the nonlinear dynamics governed by
a multidimensional cubic-quintic nonlinear Schr\"{o}dinger equation (NLSE)
received a great deal of attention. In this context, both localized vortex
solitons and nonlocalized optical vortices were studied \cite{Skarka}.
Cubic-quintic ($2+1$)-dimensional NLSE were used to study the stability of
spinning ring solitons \cite{Towers} and theoretical investigations to find
solitary solutions for the cubic-quintic ($1+1$)-dimensional NLSE have been
carried out. Dark solitary waves in the limit of small amplitude have been
found, where the NLSE was reduced to a Korteweg-de Vries equation (KdVE)
\cite{Bass}. Moreover, both algebraic solitary wave solutions \cite{Hayata}
and traveling-wave solutions \cite{Schurmann-1} have been found and
criteria for existence and stability of soliton solutions have been
established \cite{Schurmann-2}. Additionally, a theory which connects
envelope solitons of a wide class of generalized NLSEs with solitons of a
wide class of generalized KdVE have been recently carried out for arbitrary
amplitudes \cite{r5-1}; in particular, the theory was applied to find
analytical bright, gray and dark envelope soliton solutions of the
cubic-quintic NLSE and some other types of nonlinearities
\cite{r5-1}-\cite{r5-3}.

It is well known that nonlinear interactions between
intense laser beams and a plasma are responsible for
numerous nonlinear phenomena including parametric
instabilities \cite{r1}, density cavitation,
self-focusing and filamentation of light \cite{r2,r3,r4},
as well as the generation of wake fields \cite{r5}.
Intense laser beams can cause density modifications
through the ponderomotive force, enhance the electron
mass due to relativistic effects and produce electron
Joule heating. The interplay between the ponderomotive,
relativistic and Joule heating nonlinearitries has been
examined \cite{r2} in the context of laser plasma
experiments and also in ionospheric  modifications of the
Earth's ionosphere by powerful radar beams.

In this Letter, we investigate nonlinear interactions
between the light beams and the non-resonant density
perturbations, taking into account the combined effects
of the light pressure induced ion density fluctuations
and increased electron mass. We show that, under suitable
physical conditions for which our system can be described
by a ($1+1$)-dimensional cubic-quintic NLSE for the
complex electromagnetic field amplitude, bright, gray and
dark envelope solitons are analytically found. Finally, a
stability analysis is carried out, which shows that our
bright soliton solutions are stable.

\section{Basic equations}
We consider the propagation of a large amplitude
circularly polarized electromagnetic wave  with an
electric field ${\bf E} = E (\hat {\bf x} + i \hat {\bf
y})\exp(-i\omega t + i {\bf k} {\bf R})$, where $\omega$
is the wave frequency and ${\bf k}$ is the wavevector.
The light equation in the presence of electron density
perturbations in a plasma is  obtained from
\begin{equation}
\nabla  {\bf B} =\frac{4\pi}{c}{\bf J} + \frac{1}{c}\partial_t {\bf E},
\end{equation}
with
\begin{equation}
{\bf J} = -e (n_{0}+n_{1}) {\bf v}_e,
\end{equation}
\begin{equation}
{\bf B} = \nabla  {\bf A},
\end{equation}
\begin{equation}
{\bf E} = -\frac{1}{c}\partial_t {\bf A},
\end{equation}
and
\begin{equation}
\partial_t {\bf p}_e = -e {\bf E},
\end{equation}
where ${\bf B}$ is the wave magnetic field, ${\bf A}$ is the vector potential,
$n_{0}$ and $n_{1}$ are the unperturbed and perturbed electron number densities,
${\bf v}_e$ is the particle quiver velocity induced by the photons,  ${\bf p}_e
=m_e {\bf v}_e$ is the momentum, $m_e
=m_0/(1-{\bf v}_e^2/c^2)^{1/2}$ is the mass, $m_0$ is the rest electron mass,
$e$ is the  magnitude of the electron charge, and $c$ is the speed of light in
vacuum.  The perturbation of the number density $n_1$ is reinforced  by  the
light ponderomotive force.    For our purposes, we have
\begin{equation}
{\bf v}_e = \frac{e}{m_0 c}\frac{{\bf A}}{\gamma_e}
\end{equation}
in view of (4) and (5). Here we have denoted $\gamma_e =\sqrt{1+e^2 {\bf
A}^2/m_0^2c^4}$.  Combining (1), (2), (3) and (6) we obtain
\begin{equation}
\partial_t^2 {\bf A}-c^2 \nabla^2 {\bf A} + \omega_{p}^2\left(1 + N\right)
\frac{{\bf A}}{\gamma_e} = 0,
\end{equation}
where $\omega_p =(4\pi n_0e^2/m_0)^{1/2}$ is the plasma frequency,  $N =
n_{1}/n_0$, and where we have  introduced the Coulomb gauge $\nabla  {\bf A}
=0$.  We note that the term proportional to $n_1$ comes from the beating of the
light quiver velocity with the electron density  perturbations of the plasma
slow motion, while the nonlinear term in $\gamma_e$  arises due to the electron
mass increase in the light wave fields.   Supposing that ${\bf A} = {\bf A}_s
({\bf r}, \tau)\exp (i{\bf k}  {\bf R}
-i\omega t)$ + complex conjugate, where ${\bf r}$ and $\tau$ represent slowly
varying space and time coordinates, we obtain from (7)
\begin{equation}
2i\omega (\partial_{\tau} + {\bf v}_g  {\bf \nabla}_{\bf r}){\bf A}_s +c^2
\nabla_{\bf r}^2{\bf A}_s +\Omega^2 {\bf A}_s
- \frac{\omega_{p}^2(1+N){\bf A}_s}{\gamma_e}=0
 ,\end{equation}
where ${\bf v}_g ={\bf k}c^2/\omega$ is the group velocity of the light wave,
and $|\partial_\tau {\bf A}| << \omega {\bf A}$ has been invoked in view of the
WKB  approximation. We have denoted $\Omega^2 =\omega^2 -c^2k^2$.  We now derive
the equation for low-phase velocity (in comparison with the electron  thermal
speed) density perturbations that are driven by the light wave ponderomotive
force. The governing equations are the inertialess electron momentum equation
\begin{equation}
0 =e \nabla_{\bf r} \phi -m_0c^2 \nabla_{\bf r} \gamma_e -T_e \nabla_{\bf r} ln
~ (n_e/n_0),
\end{equation}
the ion continuity equation
\begin{equation}
\partial_\tau n_{i} + \nabla_{\bf r}  (n_i {\bf u}_i) =0,
\end{equation}
and
\begin{equation}
\partial_\tau {\bf u}_i +({\bf u}_i  \nabla_{\bf r}){\bf u}_i
=-\frac{e}{m_i}\nabla_{\bf r} \phi -\frac{T_i}{m_i}\nabla_{\bf r}
ln \, (n_{i}/n_0),
\end{equation}
where $\phi$ is the ambipolar potential, ${\bf u}_i$ is the fluid velocity
associated with the plasma slow motion, and $T_e (T_i)$ is the electron (ion)
temperature. The second  term on the right-hand side of (9) represents the light
pressure. Equations (9) to (11) form a closed system when the quasi-neutrality
$n_e =n_i$ is invoked. The light ponderomotive force acting on the ion fluid is
insignificant. Equation (9) shows that the electrons are  pushed away from the
region of maximum light intensity, and reinforce a space charge field  ($-\nabla
\phi$) and the associated density fluctuations.  The light ponderomotive force
is transmitted to ions through the space charge  field/ambipolar potential.
Adding (9) and (11) and letting $n_{e,i} =n_0 + n_1$,
${\bf u}_i = {\bf u}_{i0}+ {\bf u}_{i1}={\bf u}_{i1}$, we obtain for $n_1 <<
n_0$ and $|\left({\bf u}_{i1}{\cdot}\nabla_{\bf r}\right){\bf
u}_{i1}|<<|\partial_\tau {\bf u}_{i1}|$
\begin{equation}
\partial_\tau {\bf u}_{i1} =-\frac{m_0}{m_i}c^2\nabla_{\bf r}\gamma_{e1}
-\frac{C_s^2}{n_0}\nabla_{\bf r} n_1,
\end{equation}
where for consistency we have supposed $e^2{\bf A}_{s}^{2}/m_0^2c^4 <<1$ and,
consequently, introduced a small perturbation $\gamma_{e1}$ of the electron
relativistic factor $\gamma_e$ ($\gamma_e \approx 1+\gamma_{e1}$, where
 $\gamma_{e1}=e^2{\bf A}_{s}^{2}/2m_0^2c^4$) and
$C_s =[(T_e + T_i)/m_i]^{1/2}$ is the effective sound speed. According to the
above approximations, Equation (8) becomes
\begin{equation}
2i\omega (\partial_{\tau} + {\bf v}_g  {\bf \nabla}_{\bf r}){\bf A}_s +c^2
\nabla_{\bf r}^2{\bf A}_s +\left(\Omega^2-\omega_p^2\right) {\bf A}_s
- \omega_{p}^2\left[\left(N-\gamma_{e1}\right)-N\gamma_{e1}\right] {\bf A}_s=0 ,
\end{equation}
and (12), combined with the linearized version of (10), yields
\begin{equation}
\partial_\tau^2 N-
 C_s^2 \nabla_{\bf r}^2 N
= \frac{m_0}{m_i}c^2\nabla_{\bf r}^2 \gamma_{e1}.
\end{equation}
Equations (13) and (14) are the desired equations for
coherent light beams that are coupled with non-resonant
density perturbations in an electron-ion plasma. Note
that, in principle, the quantities
$\left(N-\gamma_{e1}\right)$ and $N\gamma_{e1}$, involved
in Eq. (13), could be of the same order. In the
following,  this physical circumstance will be considered
and, to this end, we seek possible stationary nonlinear
solutions of equations (13) and (14) . in the form of
envelope solitons

\section{Envelope solitons}\label{third}
We introduce ${\bf \xi}={\bf r}-{\bf V} \tau$, where
${\bf V}$ is the velocity of the nonlinear waves, and
assume ${\bf A}_s ={\bf a}(\xi) \exp(-i\Omega_0
\tau)$, where $\Omega_0$ is a constant. Hence, we readily obtain from (13) and
(14) 
\begin{equation}
2i\omega \left[\left(-{\bf V} + {\bf v}_g \right){\cdot} {\bf \nabla}_{\bf
\xi}\right]{\bf a} +c^2
\nabla_{\bf r}^2{\bf a}
+\left(\Omega^2-\omega_p^2+ 2\omega\Omega_0\right) {\bf a}
- \omega_{p}^2\left[\left(N-\gamma_{e1}\right)-N\gamma_{e1}\right] {\bf a}=0 ,
\end{equation}
and
\begin{equation}
\left({\bf V}{\cdot}\nabla_{\bf \xi}\right)^2N- C_s^2 ~\nabla_{\bf \xi}^2 N =
\frac{m_0}{m_i}c^2~\nabla_{\bf \xi}^2 \gamma_{e1}.
\end{equation}
For the sake of simplicity, we consider here the one dimensional case for which
can write $\left({\bf V}{\cdot}\nabla_{\bf
\xi}\right)^2N = V^2 \partial_{\xi}^{2}N$.
Consequently, (16) can be immediately integrated, giving \begin{equation}
N={m_0 c^2\over m_i \left(V^2
-C_{s}^{2}\right)}~\gamma_{e1} .
\end{equation}
By taking into account the explicit expression of $\gamma_{e1}$, choosing
$|{\bf V}| =|{\bf v}_g|>> C_s$, and combining (15) and (17), we easily get 
\begin{equation}
{1\over 2}\partial_{\eta}^2 {\bf \Psi}+ \lambda^2 {\bf \Psi} -
\left[q_1 |{\bf \Psi}|^2
+q_2 |{\bf \Psi}
|^4\right] {\bf  \Psi} =0,
\end{equation}
where we have introduced the following dimensionless quantities
$$
\eta = c~\xi/\omega_p,$$
$${\bf \Psi } = {e~{\bf a}\over \sqrt{2}m_0 c^2},$$
$$\lambda^2 = {\omega^2 -c^2k^2 -\omega_p^2+
2\omega\Omega_0\over 2\omega_p^2} ,$$
$$q_1 = {1\over 2}\left(\mu-1\right)$$
and
$$q_2 =-{1\over 2}\mu,$$
with $\mu = m_0 c^2 /m_i (V^2 -C_s^2)\approx m_0 c^2 /m_i
V^2$.\newline Note that, since $\mu >0$, $q_2$ is a
negative quantity. Additionally, since $N$ and
$\gamma_{e1}$ must be of the same order, from (17) is
evident that $\mu$ is of the order of the unity ($\mu
\sim 1$), but slightly greater than $1$. This circumstance is satisfied when we choose,
consistently,
 a group velocity $v_g =V\sim
\left(m_0 /m_i\right)^{1/2}c$. This justifies why we have kept both
the nonlinear terms in (18); accordingly, the terms $q_1 |{\bf \Psi}|^2$ and
$q_2 |{\bf \Psi}|^4$ are of the same order. In particular, if $\mu$ is
exactly equal to $1$ (i.e., we have exactly
$V =\left(m_0 /m_i\right)^{1/2}c$), the (18) becomes
\begin{equation}
{1\over 2}\partial_{\eta}^2 {\bf
\Psi}+ \lambda^2 {\bf \Psi} - q_2 |{\bf \Psi}|^4 {\bf  \Psi} =0, \end{equation}
which shows that a part of the relativistic mass
variation nonlinearity is exactly cancelled out by the light ponderomotive force
driven supersonic electron density  contribution.\newline If we put
\begin{equation}
{\bf \Phi}({\bf \eta},s)={\bf \Psi (\eta)}\exp \left(-i\lambda^2 s\right),
\end{equation}
where $s$ is a new dimensionless timelike variable, the (18) can be cast as
\begin{equation}
i\partial_s{\bf\Phi} + {1\over 2}\partial_{\eta}^2{\bf \Phi} -
\left[q_1 |{\bf \Phi}|^2 +q_2 |{\bf \Phi}|^4\right] {\bf\Phi} =0,
\end{equation}
Let us suppose that $\mu$ is (slightly) larger than the
unity. In this way $q_1>0$ , (21) admits bright, gray and
dark envelope solitonlike solutions. In fact, from the
results of recent investigations \cite{r5-1} that have
found a wide class of envelope solitonlike solutions of
(21), one can find, through (20), the following
solitonlike solution for $\Psi (\eta)$
\begin{equation}
\Psi (\eta)
=\sqrt{\overline{u}\left[1+\epsilon~\mbox{sech} \left(\eta
/\Delta\right)\right]} 
\exp\left\{i B \left[\frac{\eta}{\Delta}+\frac{2\epsilon}{\sqrt{1-\epsilon^2}}
~\mbox{arctan}\left(\frac{\left(\epsilon-1\right)\mbox{tanh}
\left(\eta /2\Delta\right)}{\sqrt{1-\epsilon^2}}\right)
\right]+i\phi_0\right\}~,
\label{envelope-soliton-quartic}
\end{equation}
where $\phi_0$ is an arbitrary real constant,
$\overline{u}=-3q_1/(8q_2)=3(\mu-1)/(8\mu )$, the
constants $\epsilon$, $\Delta$ and $B$ are given,
respectively, by
$$\epsilon ={\pm}\sqrt{1- {32|q
_2|V_0^2
\over3q_{1}^{2}}}~ =~{\pm}\sqrt{1- {64|\mu|
V_0^2\over 3\left(\mu -1\right)^{2}}}~~~,$$ $$\Delta
=\left(2\sqrt{2 \bigg|
-{3q_{1}^{2}\over 64~|q_2|}~+~
{V_0^2\over 2}\bigg|}\right)^{-1}~=~\left(2\sqrt{2\bigg|-{3(\mu
-1)^{2}\over 128~|\mu|}~+~ {V_0^2\over
2}
\bigg|}\right)^{-1}~~~,$$
$$
B~=V_0\Delta~~~,
$$
provided that
$$
\lambda^2~=~\frac{15 (\mu -1)^{2}}{128|\mu|}+\frac{V_{0}^{2}}{2}~~~,
$$
and the real constant $V_0$ satisfies the condition
$$-\sqrt{\frac{3(\mu -1)^{2}}{64~|\mu|}} <V_0 <
\sqrt{\frac{3(\mu -1)^{2}}{64~|\mu|}}~~~.
$$
Accordingly the definition of $\lambda^2$ implies that $$
\omega^2 -c^2k^2 + 2\omega\Omega_0 = \omega_p^2\left(1+\frac{15 (\mu
-1)^{2}}{64\mu}+ V_{0}^{2}\right)~~~,$$
which is a condition for the real constant $\Omega_0$. According to terminology
and the results of Ref.s \cite{r5-1,r5-2}, we can distinguish the following four
cases.
\bigskip
\noindent
(a). $0<\epsilon <1$ ($V_0\neq 0$): {\it up-shifted bright soliton}.\newline
$$u(\eta=0)=\overline{u}(1+\epsilon)~,~~~\mbox{and}~~~ \lim_{\eta\rightarrow
{\pm}\infty}u(\eta)=\overline{u}$$ which corresponds to a bright soliton of
maximum amplitude $(1+\epsilon)\overline{u}$ and up-shifted by the quantity
$\overline{u}$.
\bigskip
\noindent
(b). $-1<\epsilon <0$ ($V_0\neq 0$): {\it gray soliton}.\newline
$$u(\eta=0)=\overline{u}(1-\epsilon)~,~~~\mbox{and}~~~ \lim_{\eta\rightarrow
{\pm}\infty}u(\eta)=\overline{u}$$ which is a dark soliton with minimum
amplitude $(1-\epsilon)\overline{u}$ and reaching asimptotically the upper limit
$\overline{u}$.
\bigskip
\noindent
(c). $\epsilon =1$ ($V_0 = 0$): {\it upper-shifted bright soliton}.\newline
$$u(\eta=0)=2\overline{u}~,~~~\mbox{and}~~~ \lim_{\eta\rightarrow
{\pm}\infty}u(\eta)=\overline{u}$$ which corresponds to a bright soliton of
maximum amplitude $2\overline
{u}$ and u
p-shifted by the maximum quantity
$\overline{u}$.
\bigskip
\noindent
(d). $\epsilon =-1$ ($V_0 = 0$): {\it standard dark soliton}.\newline
$$u(\eta=0)= 0~,~~~\mbox{and}~~~
\lim_{\eta\rightarrow {\pm}\infty}u(\eta)=\overline{u}$$ which is a dark soliton
(zero minimum amplitude), reaching asimptotically the upper limit
$\overline{u}$. \newline Correspondingly, the (17) gives the following
solitonlike solution for the density fluctuation \begin{equation}
N(\eta)~=~\mu~|\Psi (\eta)|^2~=~{3\over 8 }\left(\mu
-1\right)\left[1+\epsilon~\mbox{sech}
\left(\eta /\Delta\right)\right]~~~. \label{density-fluctuation}
\end{equation}
On the other hand, according to Ref.s \cite{r5-1,r5-2}, the (19) has the
following bright envelope solitonlike solution:
\begin{equation}
\Psi(\eta)~=~\left[\frac{6|E_0|}{\mu}\right]^{1/4}\mbox{sech}^{1/2}~
\left[\sqrt{2|E_0|}\eta\right]\exp\left(i
\phi_0\right)~~~,
\label{mnlse-solution
}
\end{equation}
where $\phi_0$ is an arbitrary real constant and $E_0$ is
a negative real constant satisfiying the condition
$
\lambda^2 =E_0~~~.
$
The latter implies the following condition for $\Omega_0$
$
\omega^2 -c^2k^2 + 2\omega\Omega_0 + \omega_p^2\left(2|E_0|-1\right)~=~0~~~.
$
Furthermore, the (17) implies that now the density
fluctuation corresponds to the following solitonlike solution
\begin{equation}
N (\eta)~=~\mu~|\Psi
(\eta)|^2~=~\left[\frac{6|E_0|}{\mu}\right]^{1/2}\mbox{sech}~
\left[\sqrt{2|E_0|}\eta\right]~~~. \label{density-fluctuation-1}
\end{equation}
%
%
%
We now investigate the stability of plane wave solutions
of the one-dimensional equation (21). We first note that if we set $q_2=0$,
the corresponding equation is the well known defocusing cubic
Schroedinger equation which is known to be stable.
It is therefore interesting to study
if the
$q_2| \Phi|^4$ term can modify the instability.
The analysis is performed by seeking a solution corresponding
to a uniform wave train perturbed by small disturbances:
\begin{equation}
{\Phi}=[{\Phi_0}+\rho(s,\eta) ]
\exp i\{[-q_1|{\Phi_0}|^2-q_2|{\Phi_0}|^4]s+\theta(s,\eta)\},
\label{solution1}
\end{equation}
where $\rho$ and $\theta$ are considered to be small amplitude and small phase
perturbations.
We then substitute the perturbed solution in (21) and retain only the linear
terms in $\rho$ and $\theta$. Since the resulting equation is linear we can now
assume a solution for the perturbation of the form $\rho=\rho_0\exp i[K
\eta-\Omega s]$ and
$\theta=\theta_0\exp i[K \eta-\Omega s]$. The resulting dispersion relation is
the following:
\begin{equation}
\Omega^2=\frac{K^2} {4} (K^2+4q_1 |{\Phi_0}|^2+8q_2 |{\Phi_0}|^4)
\label{disp_rel}
\end{equation}
This shows that the wave train is unstable if the
perturbation $K$
lies in the range of $0<K<2|{\Phi_0}| \sqrt{-q_1 - 2q_2 |{\Phi_0}|^2}$.
According to the definition of $q_1$ and $q_2$, the instability will occur only
if $\mu>1/(1-2 |{\Phi_0}|^2)$. The maximum instability occurs at
$K=|{\Phi_0}| \sqrt{-2 q_1-4 q_2 |{\Phi_0}|^2}$.

\section{Numerical Simulations}
In this section we analyse numerically both the influence of the
quintic nonlinearity on the modulational instability and on
the stability of a class of solitonlike solutions obtained in the
previous
sections. Equation (21) is solved numerically using a
standard pseudo-spectral code with a second order Runge-Kutta
method for advancing in time. We recall that the use of pseudo-spectral code
implies the assumption of periodic boundary conditions.

\subsection{Modulational instability}
Accordingly to the linear stability analysis performed previously,
initial conditions for our numerical simulations are
 given as follows:
\begin{equation}
{\Phi}(x,0)={\Phi_0}[1 + \varepsilon Cos(L x)],
\label{in_co}
\end{equation}
where $\varepsilon$ is the amplitude of the small perturbation and is
 taken as $10^{-2}$ the amplitude of the umperturbed wave.
Without loss of generality in our simulations we have
choosen ${ \Phi_0}=1$ and $L=1$ and have considered only
one period of the perturbation. We have performed several
numerical simulations with different values of the
parameter $\mu$. For ${ \Phi_0}=1$ the theory predicts
stability for  $\mu<-1$. In Fig. 1 we show the evolution
of a plane wave in the $\eta-s$ plane for $\mu=-1.5$. The
initial wave field persists for all times. For $\mu>1$
modulational instability should occur. In Fig. 2 we show
the case of  $\mu=-0.2$; analogously with the standard
modulational instability  we observe a Fermi-Pasta-Ulam
recurrence: periodically the perturbation grows, the wave
reaches a maximum amplitude and then goes back to the
initial condition. For $\mu> 0$ a completely different
physics takes place: in Fig. 3, obatined for $\mu=0.5$,
we do not observe anymore a recurrence and as time passes
the wave amplitude increases while the its width
decreses. This phenomenon corresponds to the
initial stage of a wave collapse (see \cite{sulem}).

\subsection{Stability of solitonlike solutions}
It is well know that the cubic NLS equations ($q_2=0$)
posses solitons solutions if $q_1$ is greater than zero.
We here investigate numerically if the solitonlike
solutions described previously are stable or not. In
order to to that we simply consider a solitonlike
solution at time $s=0$ and we let evolve numerically the
eq. (21). For semplicity we restrict our analysis to a
sub class of solutions which corresponds to the case of
bright solitons with $V_0=0$. We have performed many
different numerical simulations with different values of
the parameter $\mu$. The major result obtained is the
following: if $\mu>1$ solutions are stable and for
$0<\mu<1$ are unstable. This is due to the fact that for
$\mu>1$ the coefficients in front of the cubic and
quintic nonlinearities, respectivelly $q_1$ and $q_2$,
have opposite sign and therefore there is a sort of
balance between nonlinearities that stabilizes the
solitonlike solution. This is not the case for $0<\mu<1$:
both nonlinearities have the same sign and the dispersion
is not strong enough to balance them. In Figs. 4 and 5 we
give numerical evidence of the results presented for
$\mu=1.1$ and for $\mu=0.9$ respectivelly. The first
case, Fig. 4, corresponds to a stable solutions (the wave
profile does not change as time $s$ passes). The second
case, Fig. 5, is the unstable case: a clear increase in
the wave amplitude is shown.

According to the above stability analysis we can conclude
that our soliton solutions are stable in the range where
$\mu > 1$ only; this inequality, according to section
\ref{third}, is consistent with the conditions for their
existence in our problem.

\section{Conclusions}
To summarize, we have considered the nonlinear
interaction between intense light beams and the
non-resonant density perturbations, taking into account
the relativistic  mass increase of the electrons as well
as the light beam ponderomotive force that reinforces the
density perturbations in an electron-ion plasma.  The
nonlinear coupling is governed by a pair of equations
which, in one space dimension admit stationary solutions
in the form of a planar bright and dark/gray envelope
solitons. The condition of stability of the bright
soliton-like solutions has been found numerically and it
has shown that they are stable just for the range of
parameters required in our problem.\newline
Unfortunately, the numerical stability analysis for the
family of dark and gray solitons, although revealed to be
much more difficult, is on the way. Authors expect that
the preliminary numerical analysis of the stability
carried out in the present paper will be extended to the
dark and gray solitons in a future work.

\acknowledgements
This work was partially supported by the Deutsche Forschungsgemeinschaft and
INFN Sezione di Napoli (Italy).

%

\end{document}